\documentclass[twocolumn]{elsart5p}
\usepackage{graphicx}
\usepackage{amsmath}

\def\om{\omega}
\def\be{\begin{eqnarray}}
\def\ee{\end{eqnarray}}

\def\rmF{{\rm F}}

\newcommand{\lsim}{\stackrel{\scriptstyle <}{\phantom{}_{\sim}}}
\newcommand{\gsim}{\stackrel{\scriptstyle >}{\phantom{}_{\sim}}}

\begin{document}

\begin{frontmatter}
\title{Solution of the Hyperon Puzzle within a Relativistic Mean-Field Model}
\author[MEPHI]{K.A.~Maslov},
\author[UMB]{E.E.~Kolomeitsev} \and
\author[MEPHI]{D.N.~Voskresensky}
\address[MEPHI]{ National Research Nuclear  University (MEPhI), 115409 Moscow, Russia}
\address[UMB]{Matej Bel  University, SK-97401 Banska Bystrica, Slovakia}
\begin{abstract}
The equation of state of cold baryonic matter is studied within a
relativistic mean-field model with hadron masses and coupling
constants depending on the scalar field. All hadron masses undergo
a universal scaling, whereas the couplings are scaled differently.
The appearance of hyperons in dense neutron star interiors is
accounted for, however the equation of state remains sufficiently
stiff if the reduction of the $\phi$ meson mass is included. Our
equation of state matches well the constraints known from analyses
of the astrophysical data  and particle production in heavy-ion
collisions.
\end{abstract}
\end{frontmatter}

\section{Introduction}

A nuclear equation of state (EoS) is one of the key  ingredients
in the description of neutron star (NS)
properties~\cite{Lattimer:2012nd}, supernova
explosions~\cite{Woosley} and heavy-ion
collisions~\cite{Danielewicz:2002pu,Fuchs}. A comparison of
various EoSs in how well they satisfy various empirical
constraints was undertaken in Ref.~\cite{Klahn:2006ir} for the
EoSs obtained within relativistic mean-field models (RMF) and some
more microscopic calculations and in Ref.~\cite{Dutra12} for the
Skyrme models. It turns out difficult to reconcile the constraint
on the maximum NS mass, which must be larger than $1.97\,M_\odot$
after the recent measurements reported
in~\cite{Demorest:2010bx,Antoniadis:2013pzd}, and the upper
constraints on the stiffness of the EoS extracted from the
analyses of heavy-ion collisions
(HICs)~\cite{Danielewicz:2002pu,Fuchs}.
Another relevant constraint on the EoS of the NS matter is imposed
by the direct Urca (DU)  processes, like $n\to p+e+\bar{\nu}$,
which occur as soon as the nucleon density exceeds some critical
value $n_{\rm DU}^n$. The occurrence of these very efficient
processes, even with account for the nucleon pairing, is hardly compatible with NS cooling data, if the value
of the NS mass, at which the central density becomes larger than
$n_{\rm DU}^n$, is $M_{\rm DU}^n<1.5M_{\odot}$ (the so-called
``strong" DU constraint)~\cite{Blaschke:2004vq,Klahn:2006ir}.
There should be $M_{\rm DU}^n<1.35M_{\odot}$ (the ``weak" DU
constraint)~\cite{Kolomeitsev:2004ff,Klahn:2006ir}, since
$1.35\,M_{\odot}$ is the mean value of the NS mass distribution,
as it follows from the analysis of the  observational data on NSs
in binary systems. The DU problem appears in the EoSs with linear
dependence of the symmetry energy except, maybe, most stiff
ones.  All the standard RMF EoSs and the microscopic
Dirac-Brueckner-Hartree-Fock (DBHF) EoS suffer of this linear
dependence. On the contrary, variational calculations of the
Urbana-Argonne group with A18 + $\delta v$ + UIX*
forces~\cite{APR}, as well as the RMF models with density
dependent hadron coupling constants~\cite{Typel}, generate a
weaker growth of the symmetry energy with the density, and the
problem with the DU reactions is avoided. The later models are
also able to describe NSs, as heavy as those in
Refs.~\cite{Demorest:2010bx,Antoniadis:2013pzd}.

The problems worsen if strangeness is taken into account, because
the population of new Fermi seas of hyperons leads to a softening of
the EoS and reduction of the maximum NS mass. By employing a
recently constructed hyperon-nucleon potential, the maximum masses
of NSs with hyperons are computed to be well below $1.4
M_{\odot}$~\cite{Djapo:2008au}. Also, within RMF models one
is able to explain observed massive NSs only, if one  artificially
prevents the appearance of hyperons, cf.~\cite{Djapo:2008au,Glendenning}
and the references therein. This is
called in the literature, the ``hyperon puzzle". So, the
difference between NS masses with and without hyperons proves to
be so large for reasonable hyperon fractions in the standard RMF
approach that in order to solve the puzzle one has to start with
very stiff purely nuclear EoS, that hardly agrees with the results
of the microscopically-based variational EoS~\cite{APR} and the
EoS calculated with the help of the auxiliary field diffusion
Monte Carlo method~\cite{Gandolfi:2009nq}. Such an EoS would also
be incompatible with the restrictions on the EoS stiffness
extracted from the analysis of nucleon and kaon flows in heavy-ion
collisions~\cite{Danielewicz:2002pu,Fuchs}. All suggested
explanations require additional assumptions, see discussion
in~\cite{Fortin:2014mya}. For example, the inclusion of an
interaction with a $\phi$-meson mean field, and the usage of
smaller ratiohypreon-nucleon coupling constants following the SU(3)
symmetry relations~\cite{Weissenborn:2011ut}, as well as other
modifications performed within the standard RMF approach,  all
help to increase the NS mass.

There is also another part of the ``hyperon puzzle", which attracted less attention so far. With the hyperon coupling constants introduced with the help of the SU(6) symmetry relations
the critical densities for the appearance of first hyperons prove to be rather low, $n_{\rm DU}^{H}\sim 3n_0$, cf.~\cite{KV03,Weissenborn:2011kb}. However with the appearance of the hyperons the  efficient DU reaction on hyperons, e.g. $\Lambda\to p+e+\bar{\nu}$, occurs that may potentially cause a very rapid cooling of the NSs with $M>M_{\rm DU}^H $, $M_{\rm DU}^H $ being the NS mass, at which the central density reaches the value $n^{H}_{\rm DU}$. The problem should be additionally studied with account for a suppression of hyperon concentrations and weak interaction vertices of hyperons compared to the corresponding neutron concentrations and nucleon weak interaction vertices, and with account for the hyperon pairing, cf.~\cite{Prakash,Xu:2014usa}.

In Ref.~\cite{Kolomeitsev:2004ff} two of us formulated an RMF
model, in which hadron masses and meson-baryon coupling constants
are  dependent on the $\sigma$ mean field. A working model
 MW(n.u., z=0.65) labeled in \cite{Klahn:2006ir}
as KVOR model has been constructed. This model was shown
in~\cite{Klahn:2006ir} to satisfy appropriately the majority of
experimental constraints known by that time. In
Ref.~\cite{Khvorostukhin:2006ih} the particle thermal excitations
were incorporated and the model was successfully applied to
description of heavy-ion collisions.  However  hyperons are not
included in the model.  Even without hyperons the  KVOR EoS with the
added (BPS) crust EoS from~\cite{Baym:1971pw} yields $M_{\rm
max}^{\rm KVOR}=2.01 M_{\odot}$ that fits the new
constraint~\cite{Demorest:2010bx,Antoniadis:2013pzd} only
marginally.

In the present Letter we will show that within the RMF models with hadron masses and coupling constants dependent on the $\sigma$ mean field one is able to overcome the mentioned above problems and to construct the appropriate EoS with hyperons satisfying presently known
experimental constraints.

\section{Energy density functional}

In our RMF model we include  nucleons $N=(p,n)$ and hyperons
$H=(\Lambda^0, \Sigma^{\pm,0}, \Xi^{-,0})$ interacting with mean
fields of mesons $m=\sigma, \omega, \rho,\phi$. For simplicity, we
drop the $\rho$ meson self-interaction and disregard, therefore, a
possibility of the charged $\rho$-meson condensation discussed
in~\cite{Kolomeitsev:2004ff}. As the bare masses of the particles
we take $m_N=938$~MeV, $m_\Lambda=1116$~MeV, $m_\Sigma=1193$~MeV,
$m_\Xi=1318$~MeV and neglect the small mass splitting in isospin
multiplets. Lepton masses are $m_e=0.5$~MeV and $m_\mu=105$~MeV.

Following the approach~\cite{Kolomeitsev:2004ff}, the scalar field
enters  as a dimensionless variable $ f = g_{\sigma N}
\chi_{\sigma N} (\sigma) \sigma/m_N$ and the meson-baryon coupling
constants $g_{mB}$, $B=(N,H)$, are made $\sigma$-dependent with
the help of scaling coupling constants, $ g_{mB}\chi_{m B}(f)$
with $\chi_{\omega H}(f) = \chi_{\omega N}(f)$\,, $\chi_{\rho
H}(f) = \chi_{\rho N}(f)$\,. The bare masses of hadrons $m_B$ and
$m_{m}$ are replaced in the model by the effective masses
$m_B^*=m_B\Phi_B (f)$ and $m_{m}^*=m_m\Phi_m (f)$ scaled by the
functions
\begin{align}
\Phi_N(f) & =\Phi_m (f)=1-f\,,
\nonumber\\
\Phi_H(f) & =1- x_{\sigma H} \frac{m_N}{m_H} \xi_{\sigma H} f\,,
\label{phin}
\end{align}
where $x_{\sigma H} = {g_{\sigma H}}/{g_{\sigma N}}$ and
$\xi_{\sigma H}(f)=\chi_{\sigma H}(f)/\chi_{\sigma N}(f)$.

Taking into account the equations of motion for vector fields, the
energy density of the cold infinite matter with an arbitrary
isospin  composition is recovered from the Lagrangian of the model
in the standard way, see Ref.~\cite{Kolomeitsev:2004ff}:
\begin{align}
&E[f] =  \sum_B E_{\rm kin}\big(p_{{\rm F},B}, m_B \Phi_B
(f)\big) +
\sum_{l=e,\mu} E_{\rm kin}(p_{{\rm F},l},m_l)
\nonumber \\
&\,\,+\frac{m_N^4 f^2}{2 C_\sigma^2 } \eta_\sigma(f)  +
\frac{1}{2m_N^2}\Big[ \frac{C_\om^2 \widetilde{n}_{B}^2}{\eta_\om(f)} + \frac{C_\rho^2 \widetilde{n}_{I}^2}{\eta_\rho(f)} + \frac{C_\phi^2 \widetilde{n}_{S}^2}{\eta_\phi(f)} \Big]\,,
\end{align}
\begin{align}
\widetilde{n}_{B}=\sum_B x_{\om B} n_B\,,\,
\widetilde{n}_{I}=\sum_B x_{\rho B} t_{3 B} n_B\,,\,
\widetilde{n}_{S}=\sum_{H} x_{\phi H} n_H\,, \nonumber
\end{align}
where $x_{\om(\rho) B}=g_{\om(\rho)B}/g_{\om(\rho) N}$, $x_{\phi H}=g_{\phi H}/g_{\om N}$, $g_{\phi N}=0$.

The values of the isospin projections for baryons $t_{3i}$ follow
from the Gell-Mann--Nishijima relation $t_{3B}=Q_B-(1+S_B)/2$,
where $Q_B$ and $S_B$ are the baryon electric charge and
strangeness, respectively. The baryon densities are related to the
baryon Fermi momentum as $n_B=p_{\rmF,B}^3/3\pi^2$ and the
fermionic kinetic energy density is
\begin{align}
E_{\rm kin}(p_{\rm F},m) = \int_0^{p_{\rm F}}
\frac{p^2 dp}{\pi^2} \sqrt{p^2 + m^2}
 \,. \nonumber
\end{align}
The dimensionless coupling constants are $C_m = \frac{g_{mN}
m_N}{m_m}$ for all $m$ except $\phi$. Since the ratios $x_{\phi H}$ are determined through
$g_{\om N}$, the $\phi$-field contribution enters  the energy
density with the constant $C_\phi=C_\om m_\om/m_\phi$. Here we
take $m_\om = 783\,{\rm MeV}$, $m_\phi = 1020\,{\rm MeV}$. Bare
masses and coupling constants of all mesons except $\phi$ enter the energy density
only in combinations $C_m$ and the scaling functions
$\Phi_m$ and $\chi_m$ enter only through the scaling factors
\begin{align}
\eta_m (f)=\Phi_m^2 (f)/\chi_{mN}^2 (f)\,.\label{etam}
\end{align}
Therefore we actually do not need to determine  $\Phi_m (f)$ and
$\chi_m (f)$ separately, but only $\eta_m (f)$ combinations. The
self-interaction of the scalar field introduced usually in RMF
models through a potential $U(f)$ is hidden now in the definition
of $\eta_{\sigma}(f)$. The equation of motion for the remaining
field variable $f$ follows from the minimization of the energy
density $\partial E[f]/\partial f=0\,.$ If we suppress $\phi$ and
$H$ terms, put
$\eta_\sigma=1+2\frac{C_\sigma^2}{f^2}(\frac{b}{3}f^3+\frac{c}{4}
f^4)$ and put all other scaling functions to unity, we recover the
energy density functional of the standard non-linear Walecka
$\sigma$-$\om$-$\rho$ model.

For the NS matter sustained in the $\beta$-equilibrium the Fermi
momenta of a baryon can be expressed through the baryon chemical
potential, $\mu_B$, as $p_{\rmF,
B}^2=(\mu_B-V_B)^2-m^2_B\Phi^2_B$, where $V_B\,m_N^2=C^2_{\om}x_{\om B}\tilde{n}_B/\eta_\om + C^2_{\rho}x_{\rho B}t_{3B}\tilde{n}_{I}/\eta_\rho + C^2_{\phi}x_{\phi
B}\tilde{n}_{S}/\eta_\phi$; $\mu_B$ is related to the nucleon and
electron chemical potentials as $\mu_B=\mu_n - Q_B\mu_e$. Solving
the system of equations for $p_{\rmF,B}$ and making use of the
electro-neutrality condition $\sum_B Q_B n_B=n_e+n_\mu$, where the
lepton densities are $n_e=\mu_e^3/3\pi^2$,
$n_\mu=(\mu_e^2-m_{\mu}^2)^{3/2}/3\pi^2$, we can express the
hadron densities and the total energy density through the total
baryon density $n=\sum_B n_B$. The total pressure is calculated as
$P=\sum_{i=B,l}\mu_{i} n_i-E$.

The parameters of the nucleon sector are tuned to reproduce the
properties of nuclear matter at saturation: the saturation density
$n_0$, the binding energy per nucleon $\mathcal{E}_0$, the
effective nucleon mass $m_N^*$, the compressibility modulus $K$ and the symmetry energy $\widetilde{J}$. The
coupling constants of hyperons with vector mesons are interrelated
by SU(6) symmetry relations~\cite{vanDalen:2014mqa}:
\begin{align}
&  g_{\om \Lambda} =g_{\om \Sigma}= 2g_{\om \Xi} = {\textstyle\frac{2}{3} } g_{\om N}\,, \,\,\,
 g_{\rho \Sigma} = 2 g_{\rho \Xi} = 2 g_{\rho N}\,,
\nonumber\\
& 2 g_{\phi \Lambda} = 2 g_{\phi \Sigma} = g_{\phi \Xi} = - {\textstyle\frac{2\sqrt{2}}{3} } g_{\omega N}\,, \quad g_{\rho\Lambda}=g_{\phi N} = 0 \,.
\label{gHm}
\end{align}
The coupling constants of hyperons with the scalar mean field are
constrained with the help of the hyperon binding energies per
nucleon $\mathcal{E}_{\rm bind}^{H}$ in the isospin symmetric
matter (ISM) at $n=n_0$ given by~\cite{KV03}:
\begin{align}
\mathcal{E}_{\rm bind}^{H}(n_0) &= C_{\omega}^2 m_N^{-2}x_{\omega H}n_0
- (m_N-m_N^{*} (n_0))x_{\sigma H}\,,
\label{EHbind}
\end{align}
where we suppose $\xi_{\sigma H}(f(n_0))=1$ and use
\begin{align}
&
\mathcal{E}_{\rm bind}^{\Lambda }(n_0) = -28 \,{\rm MeV},
\quad
\mathcal{E}_{\rm bind}^{\Sigma }(n_0) = 30 \,{\rm MeV},
\nonumber\\&
\mathcal{E}_{\rm bind}^{\Xi }(n_0) = -15 \,{\rm MeV}\,.
\end{align}
 The repulsive $\Sigma$ potential prevents the appearance of
$\Sigma$ hyperons in all models considered below.

The NS configuration follows from the solution of the Tolman--Oppenheimer--Volkoff
equation. For $n\lsim (0.6-0.8) n_0$, our RMF EoSs should be matched with the EoS of the NS crust, where the formation of a pasta phase  is explicitly included.
The presence of the pasta, although it may  change transport properties of the matter, affects an EoS only slightly ~\cite{MTVTC}. Therefore simplifying consideration we chose the frequently-used crust BPS EoS from~\cite{Baym:1971pw}, ignoring the pasta phase. The same BPS crust EoS was used in~\cite{Klahn:2006ir} where it was joined with the various EoSs describing the interior region.
The pressure as a function of the density for the EoSs, we consider here, intersects
the BPS pressure at density $n_1$, such that  $0.45 n_0\lsim n_1 \lsim 0.7 n_0$.
We construct the resulting pressure as a function of the density by using the BPS pressure for $n\le 0.45n_0$, then a cubic spline interpolation for $0.45 n_0 < n \le 0.7n_0$, and the pressure for beta-equilibrium matter (BEM) given by our model for $n> 0.7\, n_0$.
The choice of the interval, which includes the intersection points for both EoSs, guarantees the smoothness of the interpolation. We have also checked that a narrowing of the interpolation interval limits almost does not reflect on such observables as the star mass and radius.

\section{Models}

We discuss now two models constructed according to the principles described above.
One is the formal extension of the KVOR model from Ref.~\cite{Kolomeitsev:2004ff}
to hyperons, which we call the KVORH model now. In this example we demonstrate
the problems, which appear when one includes hyperons.
For the second model, labeled as MKVOR (or MKVORH when hyperons are included),
we propose new set of scaling functions. In Table~\ref{tab:sat-param} we present
the saturation parameters for both models and the coefficients of the expansion of
the nucleon binding energy per nucleon near the nuclear saturation density $n_0$,
\begin{align}
& \mathcal{E} = \mathcal{E}_0 +\frac{K}{2}\epsilon^2
-\frac{K^{'}}{6}\epsilon^3 +\beta^2\widetilde{J}(n) +
O(\beta^4)\,, \nonumber\\ & \widetilde{J}(n)=\widetilde{J}+
L\epsilon +\frac{K_{\rm sym}}{2}\epsilon^2+\dots\,,
\label{Eexpans}
\end{align}
in terms of  small $\epsilon=(n-n_0)/3n_0$ and $\beta=(n_n-n_p)/n$ parameters.

\begin{table}
\caption{Characteristics of KVOR and MKVOR models at saturation}
\begin{center}
\begin{tabular}{ccccccccc}
\hline\hline
\raisebox{-.23cm}[0cm][0cm]{EoS}& $\mathcal{E}_0$ & $n_0$ & $K$ & $m_N^*(n_0)$ &$\widetilde{J}$ & $L$ &$K'$ & $K_{\rm sym}$
\\ \cline{2-9}
&           [MeV] & [fm$^{-3}$] & [MeV] & $[m_N]$  & [MeV] & [MeV] &[MeV] & [MeV]
\\ \hline
KVOR & $- 16$ & 0.16 & 275 &  0.805 &  32 & 71 &  422& -86\\
\hline
MKVOR&$- 16$ & 0.16 & 240 &  0.73 &  30 & 41 &  557 & -158\\
\hline\hline
\end{tabular}
\end{center}
\label{tab:sat-param}
\end{table}

\subsection{KVORH model}

Reference~\cite{Kolomeitsev:2004ff} proposed a set of input
parameters for the RMF model, which matches the APR EoS (in the
relativistic HHJ parameterization of~\cite{HHJ}) up to $n\lsim
4n_0$, see Eq.~(61) in~\cite{Kolomeitsev:2004ff}. To fulfill the
DU constraint Ref.~\cite{Kolomeitsev:2004ff} introduced the
scaling function $\eta_\rho (f)\neq 1$, see Eq.~(63)
in~\cite{Kolomeitsev:2004ff}. Thus the model labeled as KVR in
Ref.~\cite{Klahn:2006ir} was constructed. The idea behind the KVOR
modification of the KVR model was to demonstrate that introducing
the additional scaling function $\eta_{\om}\neq 1$ one can
increase the maximum value of the NS mass without a sizable change
of the EoS for densities $n\lsim 4n_0$.  The coupling constants
for the KVOR model are given in Eq.~(58) in
Ref.~\cite{Kolomeitsev:2004ff}.  The KVOR model produces the
maximum NS mass $M_{\rm max}=2.01M_{\odot}$, and the critical
proton density for the DU reaction threshold on neutrons $n_{\rm
DU}^n=3.92n_0$ corresponding to $M_{\rm DU}^n=1.76M_{\odot}$.

In the KVORH model the parameters  $x_{\sigma H}$ deduced from
hyperon binding energies in Eq.~(\ref{EHbind}) are
\begin{align}
x_{\sigma \Lambda} = 0.599\,,\,\,\,
x_{\sigma \Sigma} = 0.282\,,\,\,\,
x_{\sigma \Xi} = 0.305\,.
\end{align}

\begin{figure}
\includegraphics[width=9cm]{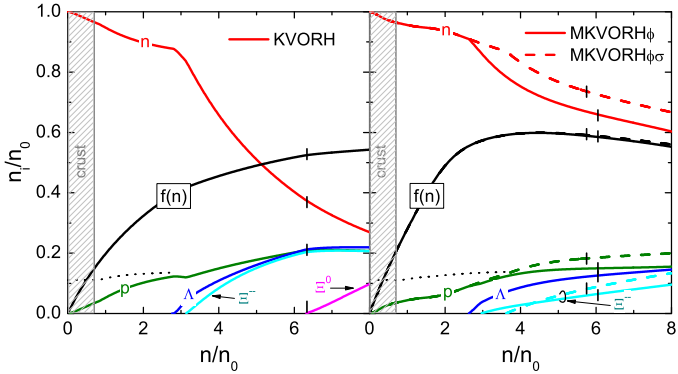}
\caption{Baryon concentrations and the variable $f$ in BEM as
functions of total baryon density for the KVORH, MKVORH$\phi$ and
MKVORH$\phi\sigma$ models. Dotted lines show the thresholds of the
DU reaction on neutrons.  For the MKVORH$\phi$ and
MKVORH$\phi\sigma$ models the thresholds are not distinguishable
on the plot scale. Short vertical bars on the lines show the
maximum densities reachable in the NS for the given model.}
\label{fig:conc}
\end{figure}

The baryon concentrations for the KVORH model are depicted in
Fig.~\ref{fig:conc} as functions of the baryon density $n$. For
$n<0.6 n_0$  the curves presented in Fig.~\ref{fig:conc} should be
replaced by those computed within the  EoS of the crust. The KVORH
model produces the maximum NS mass $M_{\rm max}=1.66M_{\odot}$.
The critical density for the appearance of first hyperons, which
is simultaneously the critical density for the onset of the DU
reactions on hyperons ($\Lambda$ in this case) is $n_{\rm
DU}^{\Lambda}=2.82n_0$, and the corresponding NS mass at which
first $\Lambda$s appear in the NS center is $M_{\rm
DU}^{\Lambda}=1.38M_{\odot}$. The total strangeness concentration
(the ratio of the number of strange quarks to the total number of
quarks) in the NS with the maximum mass is $f_{\rm S}=0.034$\,.

\subsection{MKVORH model}

Reference~\cite{Kolomeitsev:2004ff} showed that the EoS is more
sensitive to the value of $m_N^*(n_0)$ than to the compressibility
$K$. The smaller $m_N^*(n_0)$ is in a certain RMF model, the
larger is the value of the maximum NS mass. The input parameters
for the new MKVOR model are listed in Table~\ref{tab:sat-param}
together with the corresponding parameters of the nuclear binding
energy per nucleon  at saturation. We took in the MKVOR model a
smaller value of $m_N^*(n_0)$ than in the KVOR model and a smaller
value of the compressibility, $K=240$ MeV, that agrees with
canonical value $K=240\pm 20$ MeV extracted from the analysis of
giant monopole resonances (GMR)~\cite{Shlomo}.

The scaling functions of the MKVOR model are as follows:
\begin{align}
&\eta_\sigma^{-1}(f) = 1 - \frac{2}{3} C_\sigma^2 b f\big[1 +
\frac{3}{4} \big(c - \frac{8}{9} C_\sigma^2 b^2\big) f \big]
+ \frac{1}{3} d f^3 \,,
\nonumber\\
&\eta_\omega(f) = \Big(\frac{1 + z f_0}{1 + z
f}\Big)^\alpha + \frac{a_\om}{2} \left[1 + \tanh(b_\om (f -
f_\om))\right]\,,
\nonumber\\
&\eta_\rho(f) = a_\rho^{(0)} + a_\rho^{(1)} f + \frac{a_\rho^{(2)} f^2}{1 + a_\rho^{(3)} f^2}
\nonumber\\
&\quad\quad\,\, + \beta
\exp\Big[
- \gamma \frac{(f - f_\rho)^2 ({1 + e_\rho (f-f_0)^2})}
{1 + d_\rho (f-f_0) +  e_\rho (f-f_0)^2}
\Big]\,.
\label{KVORM_etar}
\end{align}
The parameters of the model and of the scaling functions are
collected in Table~\ref{tab:param}. The choice of the scaling functions is definitively not unique. Their final form was tuned to satisfy best the experimental constraints, that we demonstrate below, and to keep a connection to the KVOR model parameterization.
The first term in $\eta_\om$ is the same as in the KVOR model,
the function $\eta_\sigma$ and the first three terms in $\eta_\rho$ are basically the reparameterization of the functions of the KVOR model. The new terms, the second one in $\eta_\om$ and the last one in $\eta_\rho$ are added to control the growth of the scalar field with density.
Note that the density dependencies of our scaling functions for the $\sigma N$, $\om N$ and $\rho N$ coupling constants prove to be similar to those exploited in the DD and DD-F models with the density dependent coupling constants and bare meson field masses, cf.~\cite{Typel05,Klahn:2006ir}.

\begin{table}
\caption{Parameters of the MKVOR model}
\centering
\begin{tabular}{ccccccccccc}
\hline \hline
\multicolumn{2}{c}{$C_\sigma^2$} & \multicolumn{2}{c}{$C_\om^2$} & $C_\rho^2$ & $b \cdot 10^3$ & $c \cdot 10^3$  & $d$ & $\alpha$ & $z$ & $a_\om$
\\ \hline
\multicolumn{2}{c}{234.15} & \multicolumn{2}{c}{134.88} &  81.842\, & \,4.6750\, & $-2.9742$ & \,$-0.5$\, & 0.4  & 0.65\, & 0.11
\\ \hline
 $b_\om$ & $f_\om$& $\beta$ & $\gamma$ &$f_\rho$ &  $a_\rho$ & $a_\rho^{(1)}$ &$a_\rho^{(2)}$ & $a_\rho^{(3)}$ & $d_\rho$ & $e_\rho$
\\ \hline
 7.1\,& \,0.9\, &\, 3.11\, &\, 28.4\, & 0.522 &  0.448 & $-0.614$ & 3 & 0.8  & $-4$ & 6
\\\hline\hline
\end{tabular}
\label{tab:param}
\end{table}

The maximum NS mass increases in the MKVOR model compared to the
KVOR model  up to $M_{\rm max}=2.33\,M_\odot$. The DU threshold
values are $n_{\rm DU}^n=4.14 n_0$ and $M_{\rm DU}^n=2.22
M_{\odot}$.

In the model with hyperons (MKVORH) we use the $\sigma H$ coupling constant ratios deduced from hyperon binding energies in ISM following Eq.~(\ref{EHbind}) for $\xi_{\sigma H}=1$:
\begin{align}
x_{\sigma \Lambda} = 0.607\,, \,\,\,
x_{\sigma \Sigma} = 0.378\,, \,\,\,
x_{\sigma \Xi} = 0.307\,.
\label{sigH-MKVOR}
\end{align}
Note that the value $x_{\sigma \Lambda} \simeq 0.61$ is close to
the best value derived from hypernuclei $x_{\sigma \Lambda} \simeq
0.62$ in Ref.~\cite{vanDalen:2014mqa}.

To have an opportunity for an increase of the maximum NS mass in
the model with hyperons, we incorporate the $\phi$-meson mean
field with a scaled $\phi$ meson mass (version labeled
MKVORH$\phi$) and, additionally, allow for a scaling of the
$\sigma H$  coupling constants, $\xi_{\sigma H}(n)\neq 1$
(version MKVORH$\phi\sigma$).

In the model MKVORH$\phi$ we will exploit the very same scaling of the $\phi$-meson mass as for other hadrons $\Phi_{\phi}=1-f$. This implies the scaling function
\begin{align}
\eta_{\phi}=(1-f)^2\,.\label{etaph}
\end{align}
We use the minimal model, assuming parameterization
Eq.~(\ref{gHm}) for the vector-meson--hyperon coupling constants,
and the $\sigma H$ coupling constants from Eq.~(\ref{sigH-MKVOR}).
As the result the maximum NS mass in MKVORH$\phi$ model becomes
$M_{\rm max}=2.22\,M_{\odot}$ with the strangeness concentration
$f_{\rm S}=0.023$. The critical density for the appearance of
first hyperons is $n_{\rm DU}^\Lambda =2.63 n_0$, corresponding to
the star mass $M_{\rm DU}^{\Lambda}=1.43\,M_{\odot}$.
Although the model does not satisfy the ``strong" DU constraint
($M_{\rm DU}>1.5 M_{\odot}$), we should notice that  for reactions
on hyperons this constraint might be a bit soften since the baryon
part of the squared matrix element of the DU reaction on
$\Lambda$s is 25 times smaller than that for the DU reaction on
neutrons \cite{Prakash}, besides the pairing gaps might be not as small.

The effect of the $\xi_{\sigma H}(n)$ scaling we demonstrate at
hand of the  MKVORH$\phi\sigma$ model, where we take $\xi_{\sigma
H}(n)$ such that $\xi_{\sigma H}(n\leq n_0)=1$, and assume that
$\xi_{\sigma H}(n)$ decreases with an increase of $n$ and vanishes
for densities $n>\min\{n_{\rm DU}^H\}$. This means that
effectively we will exploit vacuum hyperon masses for
$n>\min\{n_{\rm DU}^H\}$. Note that the KVOR model extended to
high temperatures in Ref.~\cite{Khvorostukhin:2006ih} (called
there as the SHMC model) matches well the lattice data up to
$T\sim 250$ MeV provided $\sigma B$ coupling constants for all
baryons except the nucleons, are artificially suppressed, that
partially motivates our choice of suppressed values $\xi_{\sigma
H}$.

The baryon concentrations from the MKVORH$\phi$ and
MKVORH$\phi\sigma$ models are shown in Fig.~\ref{fig:conc}. The
proton fractions of the MKVORH$\phi$  and MKVORH$\phi\sigma$
models are smaller than those for the KVORH  model.  We also see
that inclusion of the $\phi$ scaling (\ref{etaph}) reduced the
hyperon population. The reduction of the $\sigma H$ coupling
constants prevents the appearance of $\Lambda$ and $\Xi^0$
hyperons and shifts the threshold density of the $\Xi^-$
appearance to higher values. Without $\Lambda$s the reaction
$\Xi^-\to \Lambda +e^-+\bar{\nu}_e$ does not occur and the DU
threshold is determined by the DU reactions on nucleons. Replacing
the values of $f(n)$ depicted in Fig.~\ref{fig:conc} for the BEM
in Eq.~(\ref{phin}) one can recover the density dependence of the
effective hadron masses and from Eqs.~(\ref{KVORM_etar}) and
(\ref{etaph}) that of the effective coupling constants.  Within
the MKVORH$\phi\sigma$ model we get $M_{\rm max}=2.29\,M_{\odot}$,
$n_{\rm DU}^{n} =3.69\,n_0$ ($M_{\rm DU}^{n}=2.09\,M_{\odot}$),
and the total strangeness concentration in the heaviest NS is
reduced to $f_{\rm S}=6.2\times 10^{-3}$.

Applying the $\phi$-mass scaling and the $\xi_{\sigma H}$ scaling
to the KVORH model we obtain for the KVORH$\phi$ model $M_{\rm
max}=1.88\,M_{\odot}$ and that the first hyperons, $\Lambda$s,
appear at the density $n_{\rm DU}^\Lambda=2.81\,n_0$  ($M_{\rm
DU}^\Lambda=1.37\,M_{\odot}$). The strangeness fraction is $f_{\rm
S}=0.035$. For the KVORH$\phi\sigma$ model we find $M_{\rm
max}=1.96\,M_\odot$, $f_{\rm S}= 9.2\times 10^{-3}$. The first among hyperons
appear  $\Xi^-$s, therefore the DU threshold is shifted to $n_{\rm
DU}^{n}=3.95\,n_0$ ($M_{\rm DU}^{n}=1.76\,M_\odot$).

Below we compare how well the EoSs obtained within the MKVORH$\phi\sigma$ and KVORH models satisfy various phenomenological constraints.

\section{Constraints on the models}

\subsection{Symmetry energy and nucleon optical potential}

\begin{figure}
\centerline{
\includegraphics[width = 9cm]{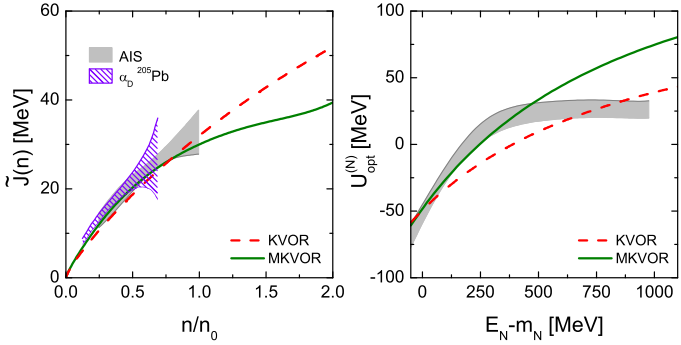}
} \caption{(Left) The symmetry energy coefficient
$\widetilde{J}(n)$ as a function of the nucleon density calculated
in the KVOR and MKVOR models. Shaded area shows the constraint
from the study of analog isobar states (AIS) in~\cite{DanielewiczLee}.
Hatched area is the constraint from the electric dipole polarizability $(\alpha_{\rm D})$ in $^{208}$Pb~\cite{ZhangChen15}.
(Right) The nucleon optical potential as a function of the nucleon
kinetic energy for ISM at $n=n_0$ for the KVOR,
cf.~\cite{Khvorostukhin:2006ih}, and MKVOR models. Shaded area
shows the extrapolation from finite nuclei to the nuclear matter
from~\cite{Feldmeier}.
 }
\label{fig:JLown}
\end{figure}

Constraints on the density dependence of the symmetry energy
[parameter $\tilde{J}(n)$ in Eq.~(\ref{Eexpans})] are  extracted
in~\cite{DanielewiczLee} from the study of the analog isobar
states and in~\cite{ZhangChen15} from the electric dipole polarizability of $^{208}$Pb nuclei. They are shown in Fig.~\ref{fig:JLown}\,(left panel) by the
shaded and hatched regions, respectively, together with the symmetry energies calculated in
the  KVOR and MKVOR models. We see that the both models follow the
lower boundary of the region.

The dependence of the nucleon optical potential on the nucleon
kinetic energy in the ISM at $n = n_0$ is shown in
Fig.~\ref{fig:JLown}\,(right panel). The shaded region is
extracted from the atomic nucleus data~\cite{Hama} and
recalculated to the case of the infinite nuclear ISM
in~\cite{Feldmeier}. The KVOR model describes the nucleon optical
potential for low and high particle energies but does not describe
it for intermediate energies. The MKVOR model describes the
nucleon optical potential rather well for nucleon energies $E_N -
m_N \lsim 400$ MeV. To fit appropriately the data at higher
particle energies, the momentum dependence of the $NN$ interaction
would be required that is not present in the mean-field approach.
The iso-vector part of the optical potential $U_{\rm opt}^n-
U_{\rm opt}^p$ is less constrained by the data, therefore we do
not show it.

\subsection{Particle flow in heavy-ion collisions}

\begin{figure}
\parbox{9cm}{\includegraphics[width = 9cm]{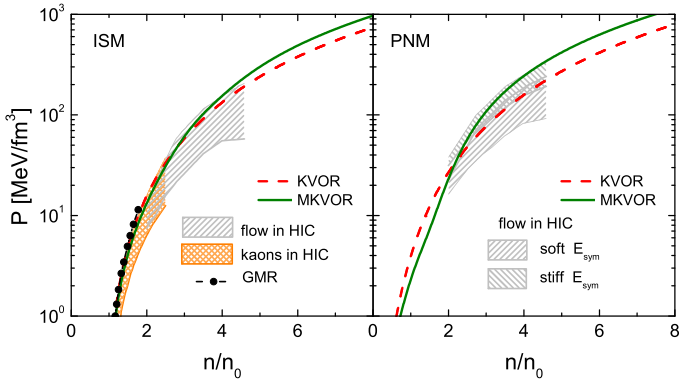} }
\caption{Pressure as a function of the nucleon density in ISM and PNM for the KVOR and MKVOR models. Hatched areas show the empirical constraints from the analyses of a particle flow in
 HICs in~\cite{Danielewicz:2002pu}, kaon production in HICs~\cite{Lynch,Fuchs}, and the GMR data~\cite{Lynch}.
}
\label{fig:Press}
\end{figure}

The analysis of the transverse and elliptical flow data in HICs
allowed~\cite{Danielewicz:2002pu} to extract a constraint on the
pressure of the ISM as a function of the nucleon density and to
reconstruct the pressure for the purely neutron matter (PNM) with
some assumptions about the density dependence  of the symmetry
energy $\tilde{J}(n)$ (soft or stiff one). Reference~\cite{Lynch} on the basis of
calculations of the kaon production in HICs ~\cite{Fuchs} provided some
restrictions on the pressure at lower densities. These constraints
are shown in Fig.~\ref{fig:Press} together  with the results of
the GMR data analysis taken from~\cite{Lynch} and the pressure
calculated in the KVOR and MKVOR models. The constraints rule out
a very stiff EoS. We see that the KVOR model satisfies the
requirements. The MKVOR model fulfills the constraints, for $n<4
n_0$ in ISM. For PNM the MKVOR curve passes through the hatched region
with stiffer $\tilde{J}(n)$, whereas the KVOR EoS is softer.
The choice of the smaller $m_N^*(n_0)$ leads to a stiffening of the EoS in ISM
and the scaling functions $\eta_\sigma (f)$ and $\eta_\om (f)$ are chosen to soften the EoS for $n\lsim 4 n_0$ to fulfill the nucleon- and kaon-flow constraints. To increase the maximum NS mass we had to stiffen the EoS in the BEM that was accomplished by the choice of the $\eta_\rho$ function.

\subsection{DU constraint}

In the BEM  the DU process on neutrons, $n \to p + e
+\bar{\nu}_e$, can occur only, if the proton fraction is high
enough so that the Fermi momenta of neutrons, protons and
electrons ($p_{{\rm F},i=n,p,e}$) satisfy the inequality $p_{{\rm
F},n} \leq p_{{\rm F},p}+p_{{\rm F},e}$. Usually, RMF models yield
uncomfortably low values of the threshold densities for these
reactions and correspondingly low values of the NS mass, $M_{\rm
DU}^n$, at which the process begins to occur in the NS center.
Every star with a mass only slightly above $M_{\rm DU}^n$ cools
down fast due to the DU process, even in the presence of nucleon
pairing, and becomes invisible for the thermal detection within
few years~\cite{Blaschke:2004vq}. Most of single NSs have
likely masses below $1.5 M_{\odot}$ in accordance with   the
type-II supernova explosion scenario~\cite{Woosley} and the
population synthesis analysis~\cite{Popov}. Therefore, it is
natural to believe that majority of the pulsars, which surface
temperatures have been measured, have masses $M\lsim  1.5
M_{\odot}$. The analysis of these data in the existing
cooling scenarios  supports the constraint $M_{\rm DU}^n\gsim 1.5
M_{\odot}$~\cite{Blaschke:2011gc,Elshamouty:2013nfa}. The adequate
description of the new data on the cooling of the
Cassiopea A also requires the absence of the DU
reactions~\cite{Blaschke:2011gc}.

In the presence of $\Lambda$ hyperons, reactions $\Lambda\to
p+e+\bar{\nu}$ may occur.  As seen in Fig.~\ref{fig:conc},
proton concentrations for the KVORH, MKVORH$\phi$ models do not
exceed the neutron-DU threshold for $n< n_{\rm DU}^\Lambda$, and for higher densities
the DU reactions on $\Lambda$s start occurring. For the
MKVORH$\phi\sigma$ model $\Lambda$s do not appear at all and the
DU processes occur for $n>n_{\rm DU}^n$.

\subsection{ Gravitational mass versus baryon mass constraint}

The unique double-neutron-star system J0737-3039  with two
millisecond pulsars provided an important constraint on the
nuclear EoS. The gravitational mass of one of the companions (B)
is very low $M_G=1.249 \pm 0.001M_{\odot}$~\cite{Kramer} which
implies a very peculiar mechanism of its creation -- a type-I
supernova of an O-Ne-Mg white dwarf driven hydrostatically
unstable by electron captures onto Mg and Ne. Knowing this
mechanism Refs.~\cite{Podsiadlowski,Kituara06} calculated the
number of baryons in the pulsar and the corresponding baryon mass:
$M_B=1.366\mbox{--}1.375M_{\odot}$~\cite{Podsiadlowski} and $M_B=
1.358\mbox{--}1.362$~\cite{Kituara06}.  The constraint of
Ref.~\cite{Podsiadlowski} can be released by
$1\%\,M_\odot$ because of a possible baryon loss and a critical
mass variation due to carbon flashes during the collapse.
Therefore one can speak about ``strong" (without the mass loss)
and ``weak" (with the mass loss) constraints on the EoS,
respectively.  Microscopically motivated EoSs, like the
relativistic DBHF EoS~\cite{FuchsDBHF}, the APR EoS~\cite{APR},
the diffusion Monte-Carlo one~\cite{Gandolfi:2009nq}, and many
RMF-based models do not fulfill the strong constraint of
Ref.~\cite{Podsiadlowski}.  Many EoSs do not satisfy the
constraint of ~\cite{Kituara06} and even the weak constraint of
Ref.~\cite{Podsiadlowski}, cf.~\cite{Klahn:2006ir}.

\begin{figure}
\includegraphics[width = 9.cm]{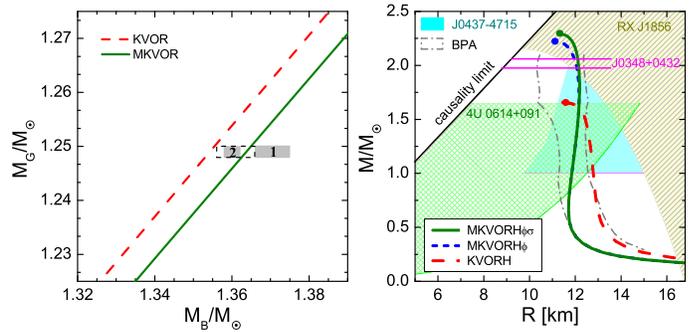}
\caption{(Left) Gravitational NS mass $M_G$ versus the baryon mass
$M_B$. Shaded rectangles show the constraints from J0737-3039(B)
pulsar derived in Refs.~\cite{Podsiadlowski} and \cite{Kituara06}
labeled by 1 and 2, respectively. The empty rectangle demonstrates
the change of the constraint~\cite{Podsiadlowski} by 1\%
$M_\odot$.  (Right) The NS mass-radius relations for the KVORH,
MKVORH$\phi$, MKVORH$\phi\sigma$ models compared with constraints
from the isolated NS RX J1856.5–3754~\cite{Trumper}, QPOs in the
LMXBs 4U 0614+09~\cite{Straaten}, the millisecond pulsar PSR
J0437-4715~\cite{Bogdanov:2012md}, and the Bayesian probability
distribution analyses (BPA)~\cite{Lattimer:2012nd}. The
horizontal lines border the uncertainty range for the mass of PSR
J0348+0432~\cite{Antoniadis:2013pzd}. } \label{fig:Pods-MR}
\end{figure}

In Fig.~\ref{fig:Pods-MR} (left panel) we plot the gravitational
NS mass $M_G$ versus the baryon mass $M_B$.  The KVOR model
matches marginally the weak constraint of
Ref.~\cite{Podsiadlowski}, whereas the MKVOR model  matches
marginally both the result from Ref.~\cite{Kituara06} and the
strong constraint of ~\cite{Podsiadlowski}, the latter was not
reproduced by the EoSs considered in Ref.~\cite{Klahn:2006ir}.
Note that hyperons do not appear in the NS with the mass of $1.25
M_{\odot}$, therefore the curves for the models KVORH and KVOR
coincide, as well as the curves for the MKVOR, MKVORH$\phi$ and
MKVORH$\phi\sigma$ models. Comparing particle concentration for
different models shown in Fig.~\ref{fig:conc} and the baryon mass
of the NS, we observe a correlation: the smaller is the proton
fraction within the density interval $n_0<n\lsim 2.5 n_0$, the
better the given EoS satisfies the baryon-mass constraint. For
$n\sim n_0$ the value of the proton fraction is correlated with
the values of $\widetilde{J}$ and $L$ in Eq.~(\ref{Eexpans}). The
value of $\widetilde{J}$ may vary only a little (from 28\,MeV to
34\,MeV or even in a narrower interval) but $L$ varies broadly in various works, e.g., cf. Fig. 4 in Ref. \cite{Lynch}. With a decrease of $L$  curves in
Fig.~\ref{fig:Pods-MR} (left panel) are shifted to the right. For
$L< 40$\,MeV (at fixed other parameters) the MKVOR curves would
pass through the shaded area 1. The smaller $L$ is for the given
EoS, the less the proton fraction is in the relevant density
interval $n_0<n\lsim 2.5 n_0$, and the better the constraint is
satisfied.

\subsection{Mass and radius constraints}

In Fig.~\ref{fig:Pods-MR} (right panel) we show mass-radius
relations of NSs for the KVORH, MKVORH$\phi$ and
MKVORH$\phi\sigma$ models. Largest precisely-measured masses of
NSs are $1.97\pm 0.04 M_{\odot}$ for PSR
J1614-2230~\cite{Demorest:2010bx} and $2.01\pm 0.04 M_{\odot}$ for
PSR J0348+0432~\cite{Antoniadis:2013pzd}. The MKVORH$\phi$ and
MKVORH$\phi\sigma$ models can describe these high-mass NSs,
whereas the KVORH model fails badly. Experimental information
about heavier NSs is plagued with large experimental errors and
with additional theoretical uncertainties, see
Ref.~\cite{Lattimer:2012nd} for a review. In
Fig.~\ref{fig:Pods-MR}  (right panel) we confront our models with
other constraints derived from the quasi-periodic oscillations in
the low-mass X-ray binary 4U~0614+091~\cite{Straaten} and thermal
spectra of the nearby isolated NS RX~J1856.5-3754~\cite{Trumper}.
More details on these constraints can be found in
Ref.~\cite{Klahn:2006ir}. In contrast to the mass determination,
there are no accurate estimates of NS radii. Some constraints were
derived recently from the X-ray spectroscopy of PSR~J0437–4715
with the proper account for the system geometry
\cite{Bogdanov:2012md} and from the Bayesian probability analysis
of several X-ray burst sources in Ref.~\cite{Lattimer:2012nd}.
These constraints are also shown in Fig.~\ref{fig:Pods-MR}. We see
that the MKVORH$\phi$ and MKVORH$\phi\sigma$  satisfy the
mass-radius constraints and produce the radii of NSs  in a narrow
interval $11.7\pm 0.5$\,km for star masses $M>0.5M_{\odot}$.

\section{Concluding remarks}

We constructed relativistic mean-field models with scaled hadron
masses and coupling constants including a $\phi$ meson mean field
and hyperons. The hyperon--vector--meson ($\omega,\rho,\phi$)
coupling constants obey the SU(6) symmetry relations. The most challenging is to fulfill the flow constraint and produce a high maximum neutron-star mass simultaneously. For that we introduced the scaling functions such that our equation of state is rather soft for
$n\lsim 4 n_0$ in the isospin symmetric matter but is sufficiently stiff in the beta-equilibrium matter. This behavior is achieved by the proper selection of the scaling functions $\eta_\om$ and $\eta_\rho$.
The inclusion of the $\phi$ meson with the mass scaled in the same
way as masses of nucleons and other mesons, but with the $\phi$ coupling
constants being fixed, allows to fulfill the empirical
constraints on the maximum neutron star mass; see curve for the
MKVORH$\phi$ model in Fig.~\ref{fig:Pods-MR} (right panel). In such an approach
the ``hyperon puzzle" can be resolved within our model without any additional assumptions, like, e.g., the change of a scheme for the choice of the hyperon--vector-meson coupling constants
from SU(6) to SU(3), as in~\cite{Weissenborn:2011ut}.
We stress that we would not succeed if we used the bare $\phi$ meson mass.
Other constraints are also satisfied except that the model
MKVORH$\phi$ produces a rather low threshold value of the neutron
star mass $M_{\rm DU}^{\Lambda}\simeq 1.44 M_{\odot}$ for the
occurrence of direct Urca reactions on $\Lambda$s. On the other hand, such a value of $M_{\rm DU}^{\Lambda}$  might be already sufficiently high to cause no problems with the too rapid cooling of neutron stars, since the DU process on $\Lambda$s might be less efficient than that on neutrons. Nevertheless we demonstrated how one can fully eliminate this possible deficiency. The model MKVORH$\phi\sigma$, where the hyperon masses do not change in
medium,  satisfies appropriately all  constraints discussed in this Letter, which are known
from the analyses of atomic nuclei, heavy-ion collisions and
neutron star data. Moreover, the maximum neutron star mass
increased. As an interesting finding, we indicate that the
smaller the proton fraction is in the density interval $n_0<n<2.5
n_0$ and the smaller the value of $L$ is at $n_0$, the better the
baryon-gravitational  mass constraint is fulfilled.

\section*{Acknowledgements}

This work was supported by the Ministry of Education and Science
of the Russian Federation (Basic part), by the Slovak Grants No.
APVV-0050-11 and VEGA-1/0469/15, and by ``NewCompStar'', COST
Action MP1304.

\end{document}